\documentclass[10pt,superscriptaddress,preprint,dvips]{revtex4}
\usepackage{epsfig}
\usepackage{graphicx}
\newcommand{\bea}{\begin{eqnarray}}
\newcommand{\eea}{\end{eqnarray}}
\newcommand{\be}{\begin{equation}}
\newcommand{\ee}{\end{equation}}
\newcommand{\ba}{\begin{array}{l}}
\newcommand{\ea}{\end{array}}

\newcommand{\half}{{\frac{1}{2}}}

\newcommand{\overleftrightarrow}[1]{\overleftarrow{#1}\hspace{-3.5mm}\overrightarrow{#1}}
\begin{document}
\title{On the $A$-dependence of nuclear generalized parton distributions}
\preprint{RUB-TPII-08/2005}
\author{V. Guzey}
\email{Vadim.Guzey@TP2.Ruhr-Uni-Bochum.de}
\affiliation{Institut f\"ur Theoretische Physik II, Ruhr-Universit\"at Bochum, D-44780 Bochum, Germany}
\author{M. Siddikov}
\email{Marat.Siddikov@TP2.Ruhr-Uni-Bochum.de}
\affiliation{Institut f\"ur Theoretische Physik II, Ruhr-Universit\"at Bochum, D-44780 Bochum, Germany}
\affiliation{Theoretical Physics Department, Uzbekistan National University, Tashkent 700174, Uzbekistan}
\begin{abstract}
We perform a microscopic evaluation of nuclear generalized parton distributions (GPDs) for spin-0 nuclei
 in the framework of the Walecka model. We demonstrate that the meson (non-nucleon)
degrees of freedom dramatically influence nuclear GPDs, which is revealed in the non-trivial
and unexpected $A$-dependence of Deeply Virtual Compton Scattering  (DVCS) observables. In particular, we find that the first moment
of the nuclear D-term, $d_A(0) \propto A^{2.26}$, which confirms the earlier prediction of
M.~Polyakov. We find that in the HERMES kinematics, contrary to the free proton case,
 the nuclear meson degrees of freedom in large nuclei enhance the nuclear
DVCS amplitude which becomes comparable to the Bethe-Heitler amplitude, and, thus,
give the non-trivial $A$-dependence to the DVCS asymmetries: as a function of the atomic number
 the beam-charge asymmetry increases
 whereas the beam-spin asymmetry decreases slowly.
 \end{abstract}
 \maketitle
\section{Introduction}
\par During the last decade generalized parton distributions (GPDs) became a standard tool
for the parameterization of the nonperturbative hadronic structure in hard processes.
Although inaccessible directly, GPDs enable us to probe the 3-dimensional structure of the
target \cite{Ralston:2001xs} and to study parton correlations in the target.
 GPDs are intensively studied both theoretically
\cite{Ji:1996nm,Ji:1997gm,Radyushkin:1996nd,Radyushkin:2000uy,Ji:1998xh,Ji:1998pc,Collins:1998be,Brodsky:2000xy,Belitsky:2001ns,Diehl:2000xz,Polyakov:1999gs,Pobylitsa:2002gw,Radyushkin:2000ge,Freund:2003qs}
and experimentally \cite{Airapetian:2001yk,Adloff:2001cn,Stepanyan:2001sm,Ellinghaus:2002zw}.
One of the key processes used for measurements of GPDs is Deeply Virtual Compton Scattering (DVCS).
\par  While most of theoretical and experimental analyses of DVCS involve the nucleon target,
there also exist ambitious projects to study DVCS on nuclei. Such experiments may provide us
with valuable information on the nuclear forces \cite{Polyakov:2002yz}
as well as on the change of nucleon properties in the nuclear medium.
 A first DVCS experiment on neon nuclei was performed at HERMES (DESY)~\cite{Ellinghaus:2002zw}.
 Another measurement of nuclear DVCS is planned  at the future Electron
Ion Collider (EIC)~\cite{Freund:2003qs,Deshpande:2002em}.
\par  By analogy with inclusive deep
 inelastic scattering (DIS) on nuclei one can expect that DVCS on nuclei will be sensitive to many nuclear phenomena
 such as shadowing, antishadowing, EMC-effect and Fermi motion.
 In addition, in the DVCS amplitude there maybe present other new nuclear
 effects which are absent in the imaginary part of the forward virtual photon-nucleon scattering amplitude
probed in inclusive DIS on nuclei.
\par DVCS on different nuclei was a subject of investigation in
\cite{Kirchner:2003wt,Freund:2003wm,Freund:2003ix,Guzey:2003jh,Freund:2002qf,
Berger:2001zb,Cano:2002tn}.
In the last two works nuclear GPDs are expressed
in terms of convolution (sum) with GPDs of separate
nucleons. This assumption is rather natural and is  based on the well-known fact that
nucleons in the nuclei are weakly bound and, thus, in hard processes
one can consider nucleus as a
collective of quasifree nucleons.
However, as we will show, this assumption is not universal,
i.e. it does not work for some of the observables.
\par
A rather interesting property of nuclear GPDs was predicted in  \cite{Polyakov:2002yz}.
Contrary to the
expectations based on the quasifree nucleon model,
the first moment of the $D$-term, which is intrinsically related to GPDs,
has nonlinear $A$-dependence, $d_A(0)\propto A^{7/3}$.
Inspired by the result of \cite{Polyakov:2002yz}, we performed a microscopic study of the nuclear GPDs
in the framework of the well-established nuclear structure model developed by Walecka and collaborators
\cite{Chin:1974sa,Serot:1997xg,Serot:1984ey}.
We confirmed the result of Ref. \cite{Polyakov:2002yz} and found
rapid $A$-dependence, $d_A(0)\propto A^{2.26}$, largely due to mesonic degrees
of freedom.
We also found that mesons significantly enhance the DVCS amplitude compared
to the nucleonic contribution.
For large nuclei in the current HERMES kinematics
\cite{Ellinghaus:2002zw}, the DVCS amplitude squared
increases as $|A_{DVCS}|^2 \propto A^{4.29}$.
We found that the nuclear asymmetries are very sensitive to the
meson  internal structure
and thus in the study of the nuclear DVCS,
one should pay particular attention to the
meson distributions in the nuclei as well as
to the quark distributions inside the mesons.
We predict that, as a function of the atomic number,
 the beam-charge asymmetry grows as $\propto A^{0.5}$.
whereas the beam-spin
asymmetry is
a slowly decreasing function of the atomic number $A$:  $A_{LU}\propto A^{-0.03}$.
 In the absence of mesonic (non-nucleonic) degrees of
freedom, the asymmetries are virtually independent of $A$.
Also, both asymmetries
should have a maximum when the DVCS and BH amplitudes have comparable magnitudes. The position of the
maximum is very sensitive to the nuclear constituents' model.
In the forward limit, our GPDs reproduce the
earlier results for nuclear light-cone distributions~\cite{Miller:2001tg,Smith:2002ci,Smith:2004dn}.
\par The paper is organized as follows.
In the Sect.~\ref{SectDterm} we demonstrate that the first moment of the
$D$-term evaluated in the Walecka model
depends on the atomic number $A$ as $d_A(0)\propto A^{2.26}$. In Sect.~\ref{SectGPD}
we evaluate the nucleonic and mesonic off-forward light-cone distributions in nuclei.
We discuss the influence of mesonic degrees of freedom on
physical observables such as  the ratio of nuclear-to-nucleon GPDs,
beam-charge and beam-spin DVCS asymmetries  in Sect.~\ref{SectProperties}.
 In Sect.~\ref{SectConclusion} we summarize our results and draw conclusions.
\section{$d_A(0)$ and energy-momentum tensor}
\label{SectDterm}
In this section we perform a microscopic calculation of the $A$-dependence
of the first moment of the $D$-term, $d_A(0)$, which is intrinsically related to nuclear GPDs.
In our analysis we use the connection  of $d_A(0)$ to the form factors of the energy-momentum tensor introduced in
\cite{Polyakov:2002yz}.
As a framework we use the field-theoretical Walecka model. In this section
nuclear constituents (nucleons and mesons) are treated as elementary pointlike objects.
The influence of their internal structure is considered in the next sections.
\par
The Lagrangian of the model in the simplest form is \cite{Serot:1984ey,Serot:1997xg}
\bea
&&{\cal L}=\bar \psi (i \hat \partial -M -g_v \hat V+g_s\phi) \psi+
\label{BasicLagrangian}\half(\partial_\mu\phi\partial^\mu\phi-
m_s^2\phi^2)-\frac{1}{4}V_{\mu\nu}V^{\mu\nu}+\frac{m^2_V}{2} V_\mu V^\mu\,,
\eea
where $V_{\mu\nu}=\partial_\mu V_\nu-\partial_\nu V_\mu;\,\,\psi$
corresponds to the field of nucleons; the  massive vector field $V$  and scalar field $\phi$
are effective fields, which represent the empirically observed dominant vector and scalar components of the $NN$ interaction.
 The relativistic MFT models based on Lagrangians of type~(\ref{BasicLagrangian}) 
 succeeded in description of such important characteristics 
as nuclear densities, the level structure of the nuclear shell model, the spin
dependence of nucleon-nucleus scattering etc. The simplest version of the model used in the work
consists of baryons and isoscalar scalar and vector mesons. A virtue of such a simple model is
that it has only two independent parameters (coupling constants) which can be fixed from the infinite nuclear matter properties. After this, all the model predictions for finite nuclei become unambiguous.
Extensions of the model with more 
independent degrees of freedom (and consequently more free parameters) can give better
quantitative description of nucleus\footnote{A study of the model dependence of the results obtained in this work is under way.}.
 Note that we neglect the pseudoscalar pion degrees of freedom since the effects
  of pions in the ground-state of spin-zero nuclei essentially average to zero due
  to the spin-dependence of the pion-nucleon coupling constant. In addition to 
  the pions, in order to achieve a truly quantitative description of nuclear properties, 
  one should ultimately include the rho-meson and electromagnetic fields. However,
  those will lead to small corrections, which we neglect in our exploratory study.
 The values of the coupling constants are fixed  from phenomenological parameters of the nuclear matter
\cite{Serot:1984ey}. 
Numerically the constants are large. Therefore,
the perturbative methods cannot be used. Instead, observing that for sufficiently
large nuclei, the nuclear density becomes large, one can use the mean-field 
approximation, when the quantum meson fields are replaced by their classical
ground-state expectation values.
The model with Lagrangian (\ref{BasicLagrangian})
is not renormalizable and should be understood
as an effective one.

Solving classical equations of motion in the nuclear rest frame~\cite{TIMORA}
and making Lorentz boosts to the infinite momentum frame,
we can obtain the light-cone wave functions necessary for evaluation of
light-cone matrix elements.
\par
As it has been shown in \cite{Ji:1996nm}, one can parameterize
the matrix element of the energy-momentum tensor
in terms of two form factors $M_A(t)$ and $d_A(t)$:
\bea\label{DT_em}
&&\langle P'|\hat T_{\mu\nu}(0)|P\rangle=M_A(t)\bar P_\mu \bar P_\nu + \frac{1}{5}d_A(t)(\Delta_\mu\Delta_\nu-g_{\mu\nu}\Delta^2)  \,,
\eea
where $\bar P=(P+P')/2, \Delta=P'-P$.
\par
Acting with the traceless operator
$$
\left(\frac{\partial}{\partial\Delta^i}\frac{\partial}{\partial\Delta^j}
-\frac{\delta_{ij}}{3}\frac{\partial}{\partial\Delta_k}\frac{\partial}{\partial\Delta^k}\right) |_{\Delta=0,\,\bar P=const}
$$
on Eq. (\ref{DT_em}), one can express $d_A(0)$ in terms of 
the matrix element of the energy-momentum tensor $\hat T_{\mu\nu}$ as
\cite{Polyakov:2002yz}
\bea
d_A(0)=-\frac{m_A}{2}\int d^3r \left(r_ir_j-\frac{\delta_{ij}}{3}r^2\right)T_{ij}(\vec r; 0),
\label{dTerm}
\eea
where $m_A$ is the mass of the nucleus;
$T_{ij}(\vec r; \Delta)$ is a shorthand notation for the Fourier of the matrix element
\bea
\frac{1}{2E}\int \frac{d^3 \Delta}{(2\pi)^3} e^{i\vec \Delta\vec r}\langle P+\Delta/2|\hat T_{ij}(0)|P-\Delta/2\rangle
\eea
 between the nuclear ground-states.
One can split the energy-momentum tensor, which can be derived from the Lagrangian
(\ref{BasicLagrangian}), into three parts
\bea
\nonumber
\hat T_{\mu\nu}=
&&\left(-V^\lambda_\mu V_{\lambda\nu}+\frac{g_{\mu\nu}}{2}V_{\lambda\rho}V^{\lambda\rho}+m_V^2V_\mu V_\nu
-g_{\mu\nu}m_V^2V_\rho V^\rho\right)+
\left(\partial_\mu \phi \partial_\nu \phi-\frac{g_{\mu\nu}}{2}(\partial_\rho\phi\partial_\rho\phi-m_s^2\phi^2)\right)\\
&&
+\left(\frac{i}{2}\left(\bar \psi (\gamma_\mu\partial_\nu+ \gamma_\nu\partial_\mu) \psi\right)
-g_V V_\mu\bar \psi\gamma_\nu \psi\right)\label{EMTensor} \,,
\eea
which will be referred to as $\hat T^V_{\mu\nu},\hat T^\phi_{\mu\nu}$ and
$\hat T^N_{\mu\nu}$, respectively.
In complete analogy with (\ref{DT_em}) one can define the form factors $M_{i/A}(t), d_{i/A}(t)$
for each of the three contributions $\hat T^i_{\mu\nu}$ to the total $\hat T_{\mu\nu}$. The form factors $M_{i/A}(t), d_{i/A}(t)$ are
connected with $M_A(t), d_A(t)$ as
\bea
&&M_A(t)=\sum_iM_{i/A}(t)=M_{N/A}(t)+M_{V/A}(t)+M_{\phi/A}(t),\nonumber\\
&&d_A(t)=\sum_id_{i/A}(t)=d_{N/A}(t)+d_{V/A}(t)+d_{\phi/A}(t).
\eea
 In Sect. \ref{SectGPD} we shall show that the form factors
$M_{i/A}(t),d_{i/A}(t)$ are closely related to the moments of GPDs of nucleons and mesons.
Assuming spherical symmetry of the considered nuclei, the final answer for $d_{i/A}(0)$ is
\bea
&&\nonumber d_{N/A}(0)=-\frac{m_A}{2}\int d^3r \left(r_ir_j-\frac{\delta_{ij}}{3}r^2\right)
\sum_n \bar \Phi_n(\vec r)\gamma_i i\partial_j\Phi_n(\vec r)-\frac{g_V}{2\,m_A}\int d^3r\,\rho_B(r)V_0(r),\\
&&\nonumber d_{\phi/A}(0)=-\frac{m_A}{3}\int d^3r\, r^2 \phi'(r)^2,\\
&&\label{dTermAll}d_{V/A}(0)=\frac{m_A}{3}\int d^3r\, r^2 V_0'(r)^2,
\eea
where the sum over $n$ in the first equation goes over the occupied levels;
$\Phi_n(r),\phi(r),V_0(r)$ are the (classical) self-consistent solutions of the
Hartree-Fock equations; $\rho_B(r)=\sum_n \Phi_n^\dagger(r)\Phi_n(r)$ is the baryon density.
\par
\begin{figure}
\includegraphics[scale=0.4]{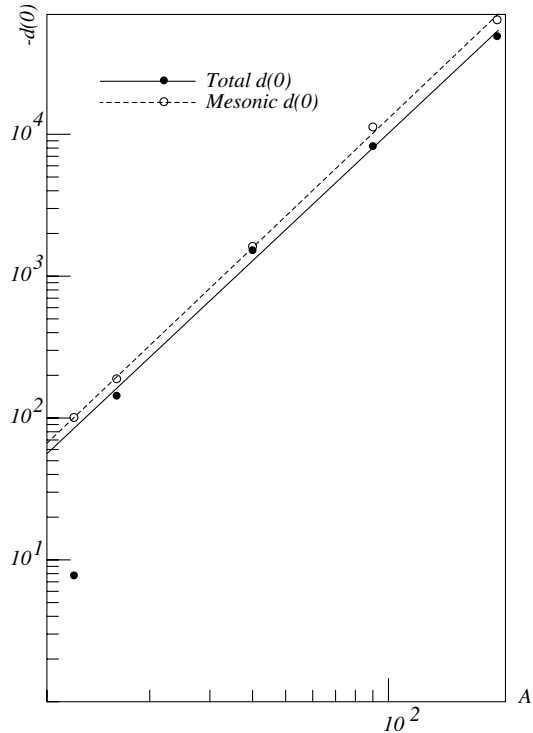}
\caption{\label{FigDTerm} $d_A(0)$ as a function of $A$. We can see that
$d_A(0)$ is mostly due to the meson contribution.}
\end{figure}

 The numerical evaluation of $d_A(0)$ for different nuclei according to Eq.~(\ref{dTermAll}) gives
the results  presented in Table~\ref{TableDTerm} and Fig.~\ref{FigDTerm}.
The most interesting feature of the obtained result is that $d_A(0)$ receives
the dominant contribution from the mesons.
Notice that $d^{mes}_A(0)=d_{\phi/A}(0)+d_{V/A}(0)$  and $d_A(0)$ for all nuclei
lie on the straight lines when plotted in the logarithmic coordinates.
The only exception is $d_A(0)$ for $^{12}C$ , where we observe
compensation of the nucleon and meson contributions. The
least-square fit to the values presented in  Table~\ref{TableDTerm} gives
\bea
d_A(0)\approx \,\,-0.308\,\, A^{2.26}.
\eea
Note that the obtained parameters contain uncertainties due to the numerical nature of
our analysis and the fitting method.
One can see that this result agrees with the estimate based on the liquid drop
model \cite{Polyakov:2002yz}
\bea
d_A(0)\propto -0.2\,A^{7/3}\,\left(1+\frac{3.8}{A^{2/3}}\right),
\eea
where the last $O(1/A^{2/3})$-term takes into account finite width of the nuclear border.
 Thus the simple power dependence  $d_A(0)\sim A^n$ exists only for relatively large-$A$ nuclei.
 In our analysis we cannot separate contributions of different parts of the nucleus. This explains why the point $^{12}C$ on the Fig.(\ref{FigDTerm}) does not lie on the straight line.
\par The $A$-dependence of the $D$-term was also examined in \cite{Liuti:2005qj},
where it was found that
\bea d_A(0)\propto A\left(1+{\cal O}(\ln A)\right).\eea
 As it is discussed in Sect.~\ref{SectGPD},
the result of \cite{Liuti:2005qj},
which is inconsistent with \cite{Polyakov:2002yz} and the result of our work, is due to the nonrelativistic approximation used by the authors.
\par
It was emphasized in \cite{Polyakov:2002yz} that the nuclear surface responsible for stability
of the liquid drop gives the dominant negative contribution to $d_A(0)$ with rapid $A$-dependence.
In the Walecka model mesons compensate
the positive contribution of the nucleons to the total pressure and, thus, provide stability
of the nucleus.
A similar phenomenon happens in the calculation of the nucleon $D$-term in
the chiral quark soliton model,
where the dominant negative contribution does also come not from
the valent quarks but from the Dirac sea \cite{Schweitzer:2002nm}.
Thus in all three models we obtain a negative value of $d_A(0)$ which comes
from ''complementary'' degrees of freedom.
\par To measure $D$-term, one should study hard processes sensitive to the real part of the
DVCS amplitude, such as the beam-charge asymmetry \cite{Kivel:2000fg}
or the DVCS total cross section in properly chosen kinematics, where the BH amplitude is suppressed \cite{Vanderhaeghen:1998uc}.
We address this issue in Sect. \ref{SectProperties} and show that the $A$-dependence of the
GPDs is observable in DVCS asymmetries.
\newline
\begin{center}
\begin{table}
\begin{tabular}{|l|r|r|}
\hline
Nucleus & $d^{mes}_A(0)$ & $d_A(0)$\\
\hline
$ ^{12}C   $ & -100.9   & -7.77   \\
$ ^{16}O   $ & -189.0   & -143.8  \\
$ ^{40}Ca  $ & -1622.5  & -1525   \\
$ ^{90}Zr  $ & -11270.2 & -8258   \\
$ ^{208}Pb $ & -64187.3 & -49195  \\
\hline
\end{tabular}
\caption{\label{TableDTerm} Values of $d_A(0)$ for different nuclei.
Contribution of mesons $d^{mes}_A(0)=d_{\phi/A}(0)+d_{V/A}(0)$ and total result $d_A(0)$.}
\end{table}
\end{center}
\section{Evaluation of nuclear GPDs}
\label{SectGPD}
\begin{figure}
\includegraphics[scale=0.4]{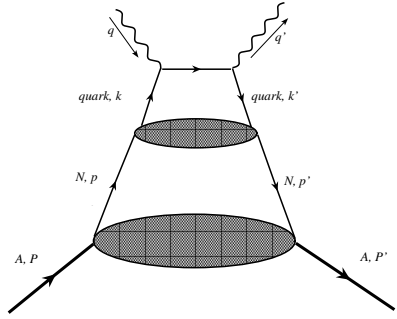}
\caption{\label{Handbag}Intuitive ''Handbag'' diagram of nuclear DVCS.}
\end{figure}

In dealing with hard processes on nuclear targets one often assumes that the
hard scattering occurs on quasifree nuclear constituents.
For the weakly-bound system it also seems quite natural to assume that one can
ignore possible ''distortion'' of the constituents in the nuclear medium
and neglect offshellness of the nucleons and mesons.
While this assumption is commonly believed to be justified for the nucleons,
we cannot estimate its accuracy for the meson degrees of freedom. However, 
since the meson GPDs are unknown even for the onshell case, 
it should be a sufficient first approximation to use the parametrization 
which only satisfies general GPDs properties.
A straightforward application of these ideas to the DVCS process
results in the convolution formula (we consider only the quark nuclear GPDs)
\bea
H_{q/A}(x,\xi,t)=\sum_i\int_{x}^1 \frac{dy}{y} H_{i/A}(y,\xi,t)
H_{q/i}\left(\frac{x}{y},\frac{\xi}{y},t\right),
\label{ConvolutionDefinition2}
\eea
where $i$ labels nuclear constituents ($\psi,V$ and $\phi$ in case of the Walecka model);
$H_{i/A}$ describe the distribution of the constituents in the nucleus;
$H_{q/i}$ are the GPDs of the free constituents.
(For a more detailed discussion, see e.g. \cite{Jaffe:1985je,Berger:1983jk}
and references therein. A generalization to the off-forward case
and non-nucleonic constituents is straightforward.)
Equation (\ref{ConvolutionDefinition2}) complies with the intuitive picture
of the nuclear hard scattering as a two-step process depicted
in Fig.~\ref{Handbag}:
the hard scattering amplitude on the nuclear target equals the hard amplitude on the free constituent
convoluted with the off-forward distribution of the constituent in the nuclear target.
Note that the convolution approximation of Eq.~(\ref{ConvolutionDefinition2}) neglects the simultaneous
coherent interaction of the virtual photon with several nuclear constituents.
Therefore, strictly speaking, Eq.~(\ref{ConvolutionDefinition2}) is applicable only for $x > 0.1$.

\par {\it Assuming} that the polynomiality conditions
\bea
\label{pol5}\int_{-1}^1x^n\,\,dx\,\,H_{i/A}\left(x,\xi,t\right)=\sum_{k=0,2,...}^{[n]}B_n^k(t)\xi^k\,,\nonumber\\
\int_{-1}^1x^n\,\,dx\,\,H_{q/i}\left(x,\xi,t\right)=\sum_{k=0,2,...}^{[n]}C_n^k(t)\xi^k\,,
\eea
fulfil separately for the constituents' GPDs in the nuclear medium and off-forward nuclear distributions,
from Eq.~(\ref{ConvolutionDefinition2}) one can obtain that
the polynomiality condition for the total GPD reads
\bea
&&\label{pol4}\int_{-1}^1x^n\,\,dx\,\,H_{q/A}\left(x,\xi,t\right)=\sum_{k=0,2,...}^{[n]}D_n^k(t)\xi^k\,,
\eea
where
\bea
D_n^{k}=\sum_{l=0}^{[n]}C_n^{l}B^{k-l}_{n-l}\,.
\label{DefinitionD_nk}
\eea
For the case of the zero-order moment, summation with
the quark charges $e_q$ of
 Eq. (\ref{DefinitionD_nk}) gives the
nuclear electromagnetic form factor
\bea
F_{em}(t)=\sum_q e_q F_{N/A}(t)F_{q/N}(t)=F_{N/A}(t)F_{N}(t),
\eea
where $F_N(t)$ is the nucleon electromagnetic form factor.
Since mesons are not charged, they do not contribute to the nuclear electromagnetic form factor.
Mathematically it follows from the antisymmetry of
the mesonic off-forward distributions
 with respect to the variable $x$, as it will be shown below.
\par
From Eq.~(\ref{DefinitionD_nk}), one obtains the first moment
\bea
&&\int^{1}_{-1}\,  x\,  dx\,  H_{q/A}(x, \xi, t)= M_{q/A}(t)+\frac{4}{5}\xi^2d_{q/A}(t)\,,
\eea
where
\bea
&&M_{q/A}(t)=\sum_i M_{q/i}(t)M_{i/A}(t)\,, \nonumber \\
&&d_{q/A}(t)=\sum_i\left(M_{q/i}(t)d_{i/A}(t)+d_{q/i}(t)\int^{1}_{-1} \frac{dy}{y}H_{i/A}(y, \xi, t)\right)\,,\label{dTermFinal}
\eea
and we introduced the conventional notation
 $M_i(t),d_i(t)$ for the form factors $B_{1}(t),C_1(t)$.
These form factors already appeared in the Lorenz-covariant expansion of the energy-momentum tensor
in the previous section.
Note that in our numerical analysis, we use a  simple parameterization of the meson GPDs,
see Eqs.~(\ref{HMesParam}) and  (\ref{OwensParam}), which corresponds to
the vanishing $d_{q/\phi}$ and $d_{q/V}$.
\par Contrary to naive expectations, as it follows from Eq.~(\ref{dTermFinal}), the total $d_A(t)$ is not reducible to
a mere sum  of the free constituents' $d_{i/A}(t)$-terms but consists of two parts.
The first term in (\ref{dTermFinal})
is a ''collective'' effect:
in  the previous section it was shown that  $d_{i/A}(0)\propto A^{2.26}$.
To evaluate the last term in  (\ref{dTermFinal}) one can use the approximation
$H_{N/A}(y,0,0)\approx A\,\delta\left(y-\frac{1}{A}\right) $ which corresponds
to an ensemble of quasifree
 nucleons and is justified for a weakly bound nucleus.
Then the last term in (\ref{dTermFinal}) is $\propto A^2$.
A direct evaluation with
the nucleon off-forward distribution
obtained in the Walecka model gives the same $A$-dependence.
Thus the predicted $A$-dependence for the total $d_A(0)$
is due to the first term in Eq.~(\ref{dTermFinal}),
which dominates in the large-$A$ limit and defines asymptotics, whereas the second term
is a
 ${\cal O}\left(A^{-0.26}\right)$-correction. Notice also that
the essential contribution to the coefficient in front of
$A^{2.26}$ comes from the mesonic degrees of freedom.
\par Now we evaluate the nuclear part of the DVCS amplitude
$H_{i/A}(x,\xi,t)$.
In complete analogy with Ref.~\cite{Ji:1998pc}
one can define off-forward distributions of nucleons and mesons in the nucleus:
\bea
&&\label{GPDNuclDef}H_{N/A}(x,\xi,t)=\half\int\frac{dz^-}{2\pi}e^{ixP^+z^-}\langle P'|\bar \psi\left(\frac{z}{2}\right)\,e^{i\int^{z/2}_{-z/2}V(\lambda)\cdot d\lambda}\gamma_+\psi\left(-\frac{z}{2}\right)|P\rangle,\nonumber \\
&&\nonumber H_{\phi/A}(x,\xi,t)=\half\int\frac{dz^-}{2\pi}e^{ixP^+z^-}\langle P'| \phi\left(\frac{z}{2}\right)i\overleftrightarrow\partial_+\phi\left(-\frac{z}{2}\right)|P\rangle, \\
&&\nonumber H_{V/A}(x,\xi,t)=\frac{1}{4x\bar P^+}\int\frac{dz^-}{2\pi}e^{ixP^+z^-}\langle P'|V_{+\alpha}\left(\frac{z}{2}\right)V^{\alpha}_+\left(-\frac{z}{2}\right)|P\rangle\\
&&+\frac{m_V^2}{4 x\bar P^+}\int\frac{dz^-}{2\pi}e^{ixP^+z^-}\langle P'| V_+\left(\frac{z}{2}\right)V_+\left(-\frac{z}{2}\right)|P\rangle\,.\label{HVdef}
\eea
The definitions of the nucleon and meson GPDs in Eq.~(\ref{SumRule}) are chosen such that
there is one-to-one correspondence to the $++$ component of the energy momentum 
tensor of the model, see Eq.~\ref{EMTensor}. This automatically guarantees conservation of
the momentum sum rule
\bea
\sum_i \int_{-1}^1 dx\,x\,H_{i/A}(x,0,0)=1.
\eea
\par Notice that the large-$A$ nucleus in the Mean Field Theory approach
in several respects resembles the nucleon in the large-$N_c$
chiral quark soliton model. A large number of particles in both models enables us to represent
the interaction of a single particle with the ensemble of other particles as the interaction with the central potential.
Forward distributions of
nucleons (quarks)
 in both cases are strongly peaked at $1/A\,\,(1/N_c)$. So we expect that
the off-forward distributions should have similar qualitative features.
\par Using standard steps \cite{Diehl:2000xz,Petrov:2002jr}, see also
the Appendix
for more details, we can obtain from Eq.~(\ref{GPDNuclDef})
\bea
&&H_{N/A}(x, \xi, t)=    \label{HNA}
\frac{m_A}{2\pi}\int\frac{dz^-}{2\pi}e^{ix\bar P^+ z^-}\int d^3 X \sum_n\bar\Phi_n\left(\frac{z}{2}-\vec X\right)P\,e^{i\int^{z/2}_{-z/2}V(\lambda)\cdot d\lambda}\gamma_+ \Phi_n\left(-\frac{z}{2}-\vec X\right).
\eea
To simplify evaluations, numerically it seems reasonable to use the approximation $V_\mu(x)\approx \delta_{\mu 0}\bar V$
in the path exponent since actually we have "cutoff" due to the wave functions $\Phi_n(r)$.
We define the average value $\bar V$ as
\bea
\bar V\simeq\frac{1}{A}\int d^3X\,V_+(X) \rho_B(X) \,,
\eea
which ensures that the approximate and exact formulas give the same results for the
first moment of $H_{N/A}$ (conservation of the nuclear momentum sum rule).
This approximation implies that the Wilson link just results in a shift of the whole parton distribution
by the value $\delta x\propto \bar V$.
The evaluation
 with the exact expression (\ref{HNA}) shows that Wilson links results not only in the
shift but also in the broadening of the nucleon
distribution.
 However such ''broadening''
is negligible, especially for large nuclei.
The expressions for off-forward distributions look more compact in the momentum representation:
\bea
&&\nonumber H_{N/A}(x, \xi, t)=\\
&&\int \frac{d\bar p^+d^2\bar p_\perp}{(2\pi)^3}
\delta\left(x+\delta x-\frac{\bar p^+}{\bar P^+}\right)
\sum_n
\bar \Phi_n\left((x-\xi),\bar p_\perp+
\frac{\Delta_\perp}{2}\right) \gamma_+
\Phi_n\left((x+\xi),\bar p_\perp-\frac{\Delta_\perp}{2}\right)
\,,\nonumber\\
&&\nonumber H_{\phi/A}(x, \xi, t)=\\
&&\int \frac{d\bar p^+d^2\bar p_\perp}{(2\pi)^3}
\delta\left(x-\frac{\bar p^+}{\bar P^+}\right)\bar p^+
\phi\left((x-\xi),\bar p_\perp+
\frac{\Delta_\perp}{2}\right)
\phi\left((x+\xi),\bar p_\perp-\frac{\Delta_\perp}{2}\right)
\,,\nonumber\\
&&\nonumber H_{V/A}(x, \xi, t)=\\
&&\int \frac{d\bar p^+d^2\bar p_\perp}{(2\pi)^3}
\delta\left(x-\frac{\bar p^+}{\bar P^+}\right)\frac{\bar p_\perp^2-\Delta_\perp^2/4+m_V^2}{4x\bar P^+}
V_+\left((x-\xi),\bar p_\perp+
\frac{\Delta_\perp}{2}\right)
V_+\left((x+\xi),\bar p_\perp-\frac{\Delta_\perp}{2}\right) \,.
\eea
\par We can see that similarly to the gluonic GPD, the vector off-forward distribution
 is singular at $x=0$.
Using explicit expressions for the off-forward distributions, one can check that
they fulfil required sum rules. For instance, the momentum sum rule reads:
\bea
&&\nonumber\int x\,dx\,H_{N/A}(x,0,0)+\int x\,dx\,H_{\phi/A}(x,0,0)+\int x\,dx\,H_{V/A}(x,0,0)=\\
&&=\frac{1}{\bar P^+}\int d^3X \left(-V_{+\lambda}(X)V_{+}^{\lambda}(X)+
m_V^2 V_+(X)V_+(X)+\partial_+\phi(X)\partial_+\phi(X)\right.\nonumber\\
&&\label{SumRule}\left.+\bar \psi(X)\gamma_+i\partial_+\psi(X)-g_V\bar\psi(X)\gamma_+V_+(X)\psi(X)\right)=
\frac{\int d^3 X\,\, \hat T^{++}}{P^+}= 1 \,,
\eea
where $\hat T^{++}$ is the $++$-component of the energy-momentum tensor
and the integral is performed over the hypersurface $X_+=const$ \cite{Serot:1984ey}.
One can easily check that the coefficients in front of $\xi^2$ in the first moments
$\int x\, dx\,H_{i/A}(x,\xi,t)$  of the off-forward distributions coincide with the results obtained in the previous section
from the energy-momentum tensor form factors, see Eq.~(\ref{dTermAll}).
\par
\begin{figure}[h]
\hspace{-1.5cm}
\includegraphics[scale=0.3]{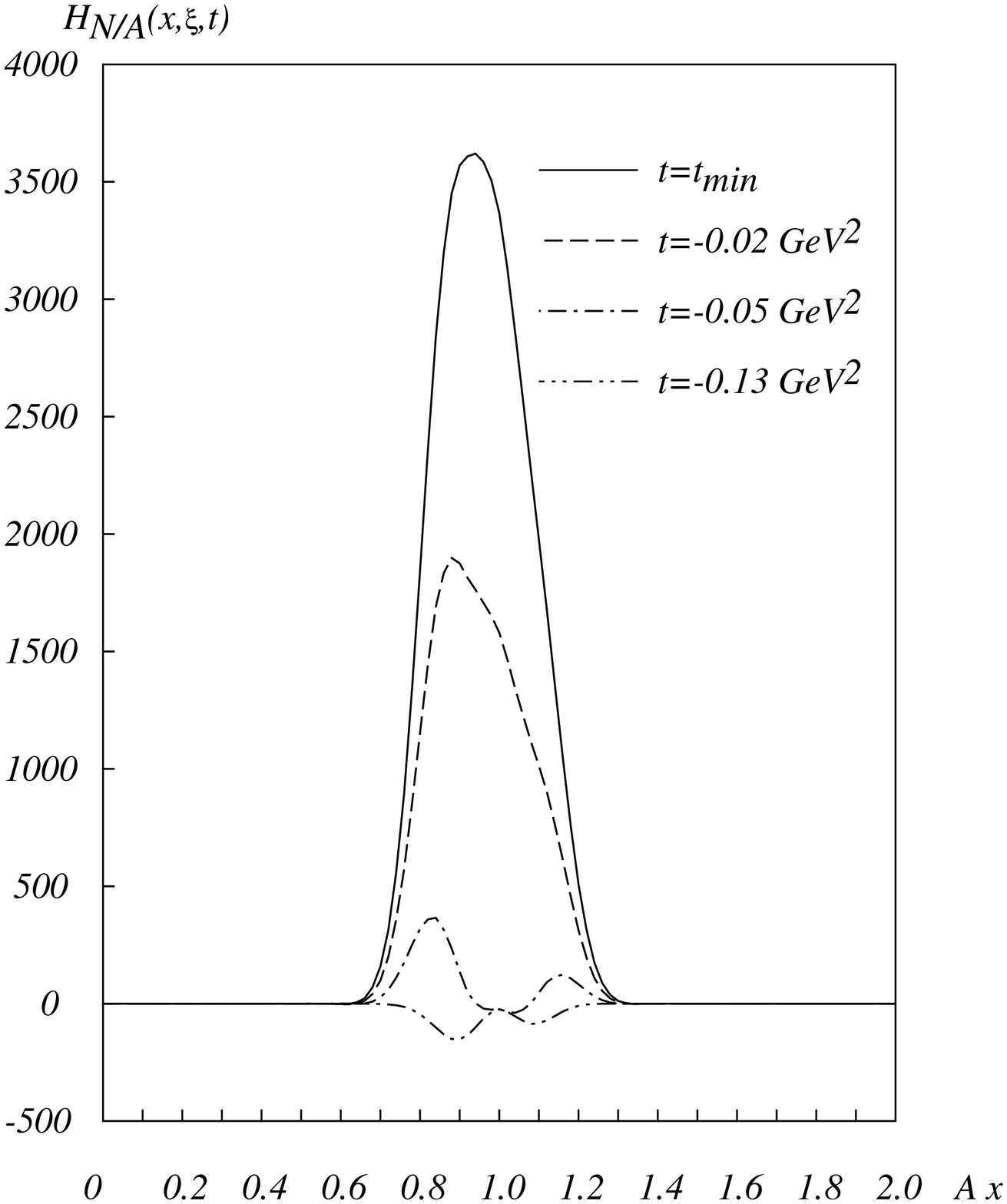}\hspace{-1.2cm}
\includegraphics[scale=0.3]{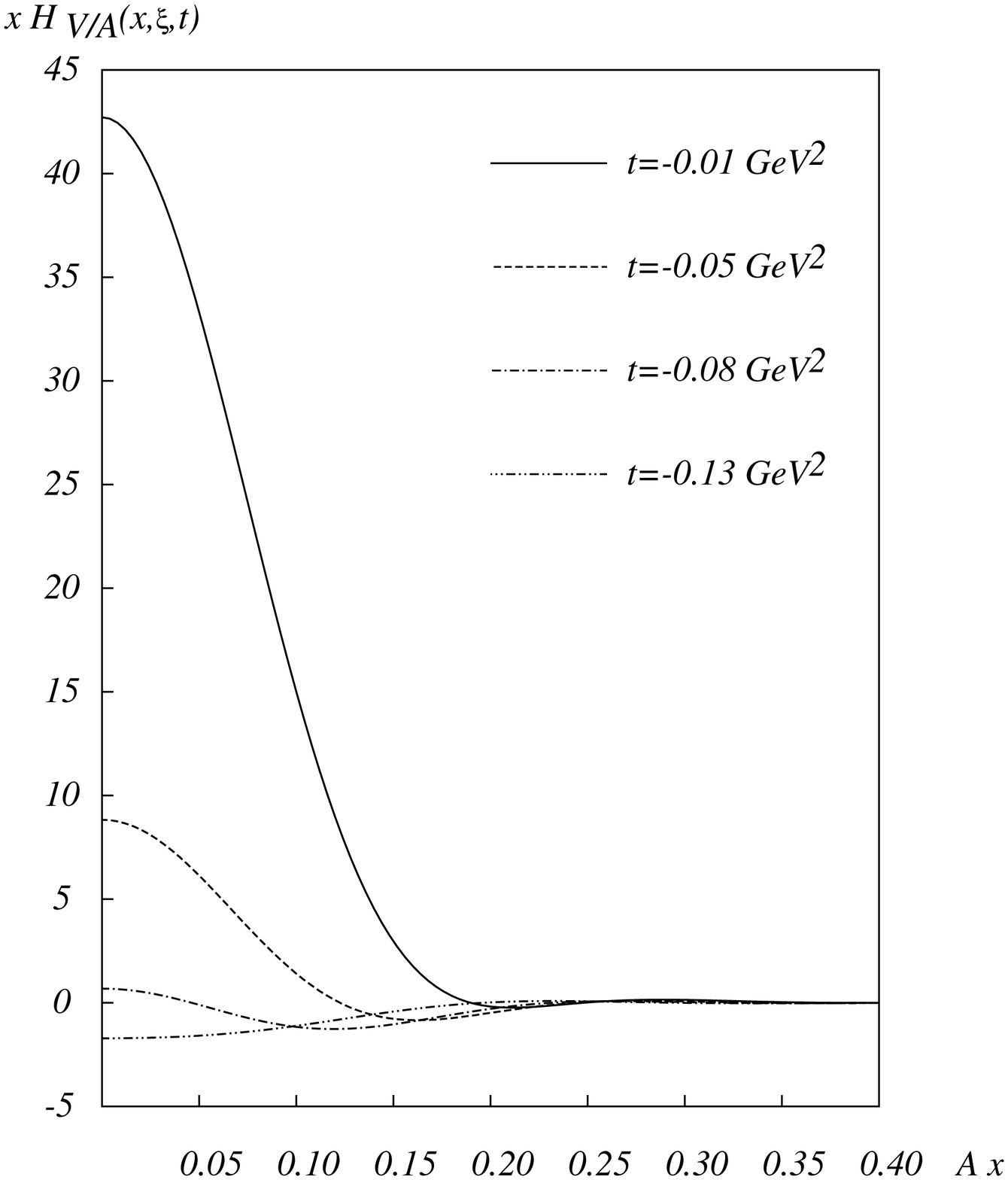}\hspace{-1.2cm}
\includegraphics[scale=0.3]{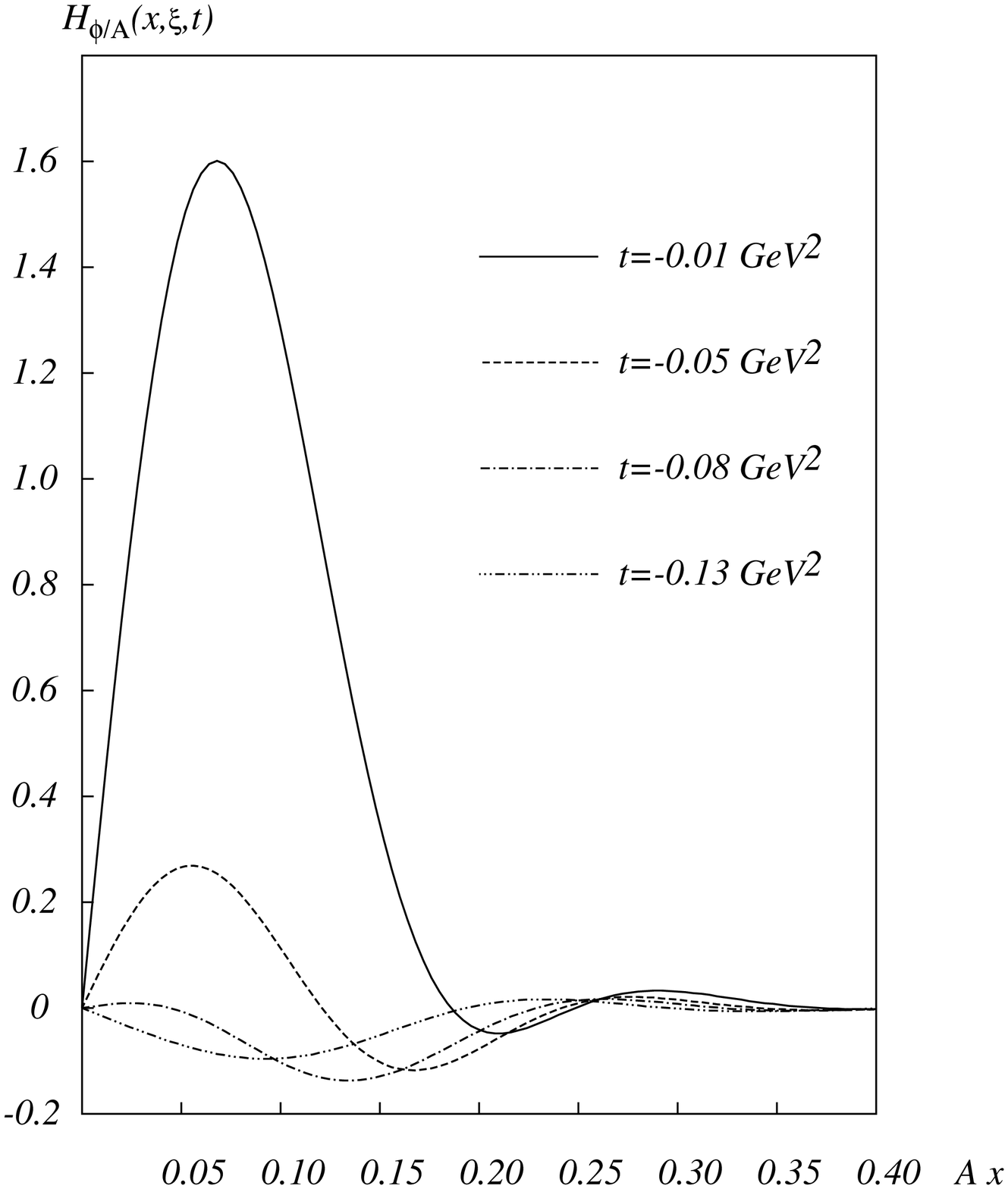}
\caption{\label{partondens}
Off-forward distributions of nuclear constituents for $^{40}Ca$
at fixed $\xi=\xi_{HERMES}\approx 0.045$ and different $t$;
$t_{min}=-4\xi^2m_A^2/(1-\xi^2)$.}
\end{figure}

 Plots of the off-forward distributions $H_{i/A}$ as functions of
the variable $A\,x$ at fixed values
of the other parameters are given in Fig.~\ref{partondens}.
We can see that all the off-forward distributions have a very strong $t$-dependence, which is similar
to the $t$-dependence of the nuclear form factor
$F_{N/A}(t)$.
The nucleonic off-forward distribution (the leftmost panel of Fig.~\ref{partondens}) has a pronounced maximum at the point $A\,x\approx 1$.
\section{Predictions for physical observables}
\label{SectProperties}
To make predictions for physical observables, we should use a
model describing the internal structure of the constituents,
which is parametrized
 by $H_{q/i}(x,\xi,t)$ in Eq.~(\ref{ConvolutionDefinition2}).
 While the resulting GPDs $H^{q/A}(x,\xi,t)$ as well as all the quantities including them
 (cross sections, asymmetries etc.) are very sensitive to the
details of the model used to parameterize $H_{q/i}(x,\xi,t)$, we expect that the ratios
of the nuclear-to-nucleon quantities should be less model dependent.
Since modelling of the  nucleon GPDs (especially in external meson fields) is beyond the scope of the current work,
we use the simplest known parameterization - the double distributions \cite{Radyushkin:2000uy}
supplemented with the $D$-term~\cite{Kivel:2000fg}
\bea
&&\nonumber H_{N}(x,\xi,t)\equiv \sum e_q^2\, H_{q/N}=F_N(t)\int_{-1}^1 d\beta \int_{-1+|\beta|}^{1-|\beta|}d\alpha
\delta(x-\beta-\alpha\xi)h(\beta,\alpha)q_{N}(\beta)+
\theta\left(1-\frac{x^2}{\xi^2}\right)D_N\left(\frac{x}{\xi},t\right),
\eea
where $q_{N}(x)\equiv\sum_q e_q^2\, q_{q/N}(x)$ is a standard flavour combination measured in DIS;
$h(\beta,\alpha)=\frac{3}{4}\frac{(1-|\beta|)^2-\alpha^2}{(1-|\beta|)^3}$~\cite{Radyushkin:2000uy};
$D_{N}(z,t)=-\sum_q e_q^2 /N_f \, 4\,(1-z^2)C^{3/2}_1(z)$~\cite{Kivel:2000fg} (we used $N_f=2$
and neglected the $D^u-D^d$ difference,which is formally suppressed
by the $1/N_c$ factor).
For the parton distributions  $q_{q/N}(x)$ we used
the CTEQ5L parameterization \cite{Lai:1999wy}.
\par Since
to the best of our knowledge, the fit for quark distributions in mesons
was done only for pions~\cite{Owens:1984zj} but not for mesons we need,
we used a simple model for the mesonic GPDs
\bea
H_{V}(x,\xi,t)=H_{\phi}(x,\xi,t)=H_{mes}(x,\xi,t)\equiv
\sum_q e_q^2\,H_{q/mes}(x,\xi,t)=\frac{5}{18}
\left(\delta\left(x-\half\right)-\delta\left(x+\half\right)\right) \,.
\label{HMesParam}\eea
This model corresponds to  a GPD of the weakly bound quark-antiquark pair.
We used $N_f=2$ and the same parameterization for both flavours since the considered mesons have
isospin zero.
 The parameterization (\ref{HMesParam}) has correct symmetry properties,
which can be derived from the $C$-parity, and satisfies the momentum sum rule
\bea
\int_{-1}^1 \sum_{q} x\,dx\, H_{q/mes}(x,0,0)=1.
\eea
We have neglected possible $\xi$-dependence for the {\it meson} GPDs $H_{mes}(x,\xi,t)$ since
in the kinematics of nuclear DVCS
 $\xi \ll 1$, $t\langle r_{mes}^2 \rangle\ll 1$,
and give only small corrections compared to the forward case.
Notice that this is not true for the nuclear form factors since the radii of the considered
nuclei are $\langle r^2\rangle^{1/2}\sim 3\div 5\, fm$ and, thus, values $t\sim 0.1\,GeV^2$ are not small
and the $t$-dependence cannot be neglected.
 For the sake of comparison we have also investigated
how our predictions change if for $H_{V}$ and $H_{\phi}$, we use a different model,
\bea
H_{q/mes}(x,\xi,t)=q_{\pi}(x)-q_{\pi}(-x) \,, \label{OwensParam}
\eea
where $q_{\pi}$ are the pion PDFs parameterized by Owens~\cite{Owens:1984zj}.
\par
\begin{figure}
\includegraphics[scale=0.3]{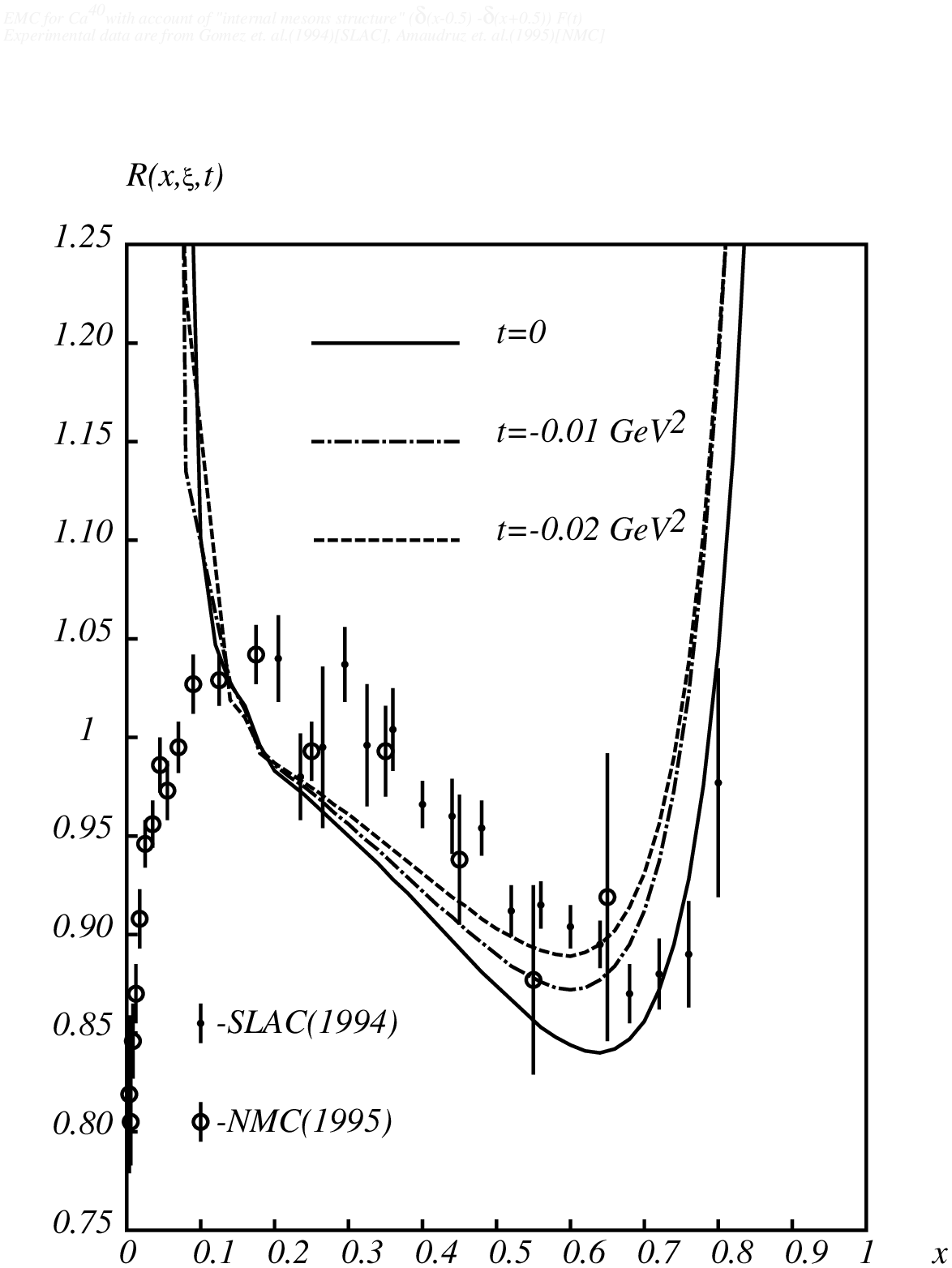}
\includegraphics[scale=0.3]{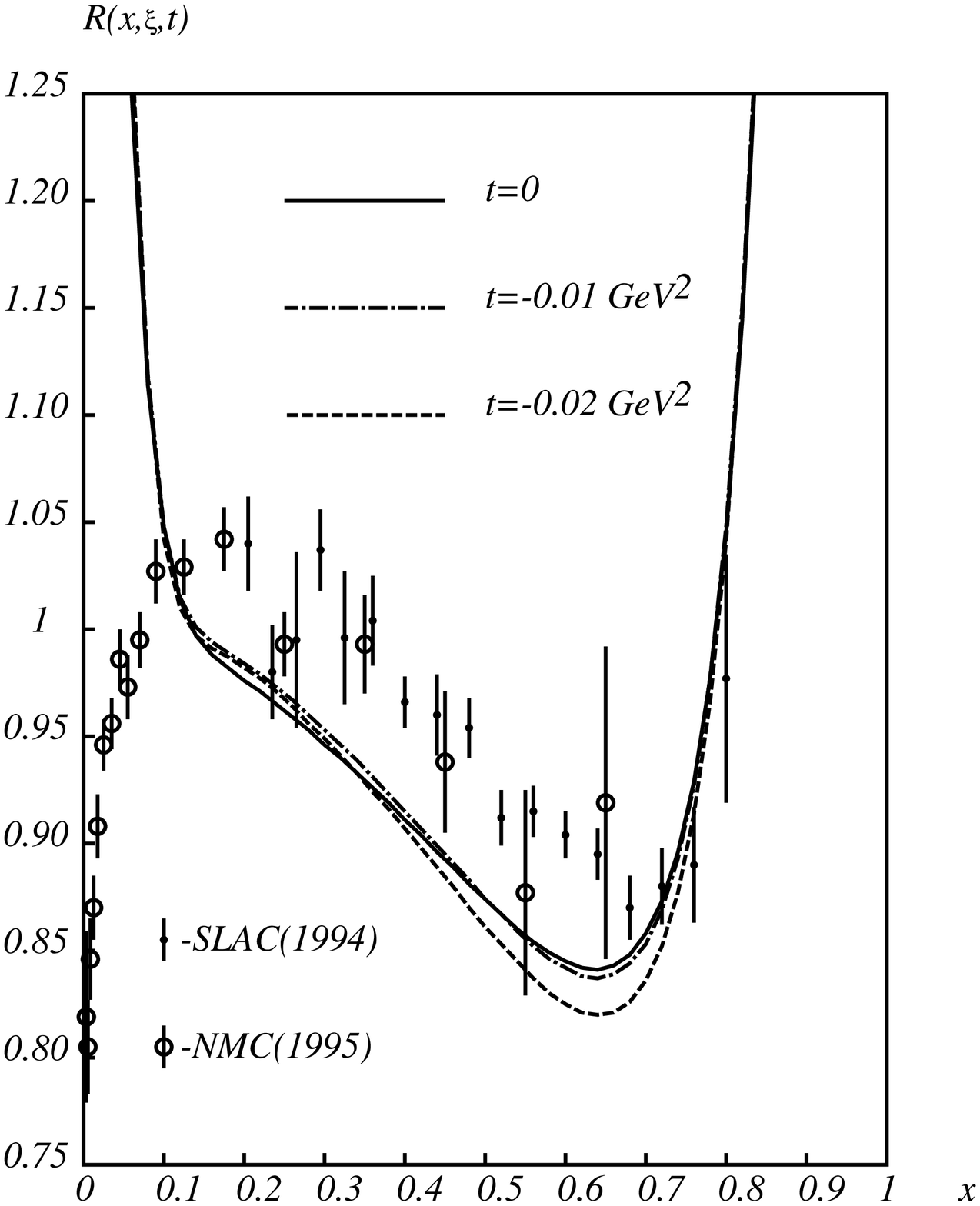}
\caption{\label{EMC}Typical behaviour of the function
$R(x,\xi,t)$ (off-forward EMC effect) for $^{40}Ca$ and different values
of  $t$. The left panel corresponds to the parameterization (\ref{HMesParam}), the right panel corresponds
to the  parameterization~(\ref{OwensParam}).
Experimental data are from SLAC~\cite{Gomez:1993ri} and NMC~\cite{Amaudruz:1995tq}
for the forward case.}
\end{figure}

 The first quantity that we consider is the ratio
\bea
R(x,\xi,t)=\frac{\sum_q e_q^2 H_{q/A}(x_A,\xi,t)}{F_{N/A}(t) H_{N}(x,\xi,t)},
\label{OffEMC}
\eea
which in the forward limit reduces to the ratio of the structure functions (to the leading order in $\alpha_S$)
\bea
R(x)=\frac{F_{2\,A}(x_A)}{A\,F_{2\,N}(x)} \,,
\eea
where $x=A\,x_A$.
The dependence of the function $R(x,\xi,t)$ on $x$
at fixed $x_{Bj}=0.09\,(HERMES)$ and different values of $t$ is given in
Fig.~\ref{EMC}.
For comparison we also present
the SLAC \cite{Gomez:1993ri} and NMC \cite{Amaudruz:1995tq} data for the ratio
$R(x)$.
Notice that the meson contribution enhances $R(x,\xi,t)$ at small $x$ above the
pure nucleonic contribution previously obtained in \cite{Miller:2001tg, Smith:2002ci}.
It is important to emphasize that the convolution approximation, which implies the interaction
with the nuclear constituents one at a time, ignores the physics of the simultaneous
coherent interaction with several constituents, which becomes important for $x < 0.1$,
 see e.g.~\cite{Piller:1999wx}. This explains
the fact that our calculation gives $R(x) > 1$ for $x < 0.1$, in dramatic contrast to the data.
While our calculations provide a fairly good description of $R(x)$ for $x > 0.1$, our model predicts
a significant enhancement of the nuclear anti-quark distributions, which contradicts the
experiment~\cite{Alde:1990im}. This problem is typical for any model of nuclear structure, which
involves a significant fraction of non-nucleonic meson degrees of freedom~\cite{Frankfurt:1988nt}.
\par The difference between
$H^{q/A}(x,\xi,t)$ and $H^{q/N}(x,\xi,t)$
can be also probed through the
DVCS beam-charge and beam-spin asymmetries
\bea
A_C(\phi)=\frac{\sigma^+(\phi)-\sigma^-(\phi)}{\sigma^+(\phi)+\sigma^-(\phi)}, &&
A_{LU}(\phi)=\frac{\overrightarrow\sigma(\phi)-\overleftarrow\sigma(\phi)}{\overrightarrow\sigma(\phi)+\overleftarrow\sigma(\phi)},
\eea
where $\phi$ is the angle between
the lepton and hadron scattering planes;
 $\sigma^\pm$ and $\overrightarrow\sigma,
\overleftarrow\sigma$ denote cross-sections of unpolarized electron/positron
and longitudinally polarized leptons,
respectively~\cite{Belitsky:2001ns}.
The asymmetries $A_C(\phi),A_{LU}(\phi)$ directly measure interference between Bethe-Heitler and DVCS amplitudes.
We perform
the evaluation of the beam-charge and beam-spin asymmetries in the kinematics
similar to that used at HERMES~\cite{Ellinghaus:2002zw}:
 $\langle x_{Bj}\rangle_{per\,nucleon}=0.09,
\langle Q^2\rangle=2.2\,GeV^2, \langle t\rangle=-0.01\,\, GeV^2$.
In this kinematics due to the small value of the prefactor $-t/4\,m_A^2$, one can
safely ignore
the contributions of the GPD $E(x,\xi,t)$ and magnetic form factor $F_2(t)$
compared to those of $H(x,\xi,t)$ and $F_1(t)$, see Ref.~\cite{Belitsky:2001ns} for more details and a
complete set of formulas describing DVCS process both for unpolarized and polarized cases.
\par In HERMES kinematics, $1/Q$ is a modest expansion parameter
for the nuclear hard scattering and one should check if the
expansion really works. For instance, due to mesons the nuclear DVCS amplitude
squared is {\it NOT} negligible compared to the Bethe-Heitler amplitude squared
(in the denominator of the expression for the DVCS asymmetries),
though numerically it is suppressed by the factor $t/Q^2$.
\par Using our results for nuclear off-forward distributions, we calculate
the dominant (leading-twist) harmonics
of the beam-charge and beam-spin harmonics,
\bea
A^{cos}_C=\frac{1}{\pi}\int_0^{2\pi} d\phi\,\cos \phi\, A_C(\phi);
&&A^{sin}_{LU}=\frac{1}{\pi}\int_0^{2\pi} d\phi\,\sin \phi\, A_{LU}(\phi) \,.
\label{AssymLTw}
\eea
In order to study the role of the mesons, we give the answer for the full calculation and
for the calculation, where the contribution of the $\phi$ and $V$ mesons was neglected.
The results are summarized in Table~{\ref{tabasym}} in terms of the ratios of the
nuclear to the free proton asymmetries. The second and third columns correspond to the
calculation without the nuclear mesons; the fourth and fifth columns correspond to the full
calculation with the meson GPDs parameterized by Eq.~(\ref{HMesParam});
the sixth and seventh  columns correspond to the full
calculation with the meson GPDs parameterized by Eq.~(\ref{OwensParam}).
\par
One can see from Table~{\ref{tabasym}} that, in the absence of the meson contributions,
 both asymmetries are practically independent of the atomic number. In this case
the DVCS amplitude squared is
small compared to the BH amplitude squared for all nuclei.
\par
In the presence of mesons
(the full calculation) we can see that
beam-charge asymmetry is a {\it growing} function of
the atomic number whereas the beam-spin asymmetry is 
weakly-dependent of  the atomic number;
the DVCS amplitude {\it increases}.
A least-square fit gives the following approximate $A$-dependence:
$A^{cos}_{C\,\,A}/A^{cos}_{C\,\,N}\propto A^{0.5}$;
$A^{sin}_{LU\,\,A}/A^{sin}_{LU\,\,N}\propto A^{-0.03}$  for all $A$;
 the ratio of the DVCS  amplitudes
 squared $|A_{DVCS\,A}/A_{DVCS\,N}|^2\propto A^{4.29}$.
 The natural explanation of such
a behaviour is that the
asymmetries are large when both DVCS and BH amplitudes have comparable magnitudes and decrease
when one of the amplitudes essentially overcomes the other.
Thus in the considered kinematics
the asymmetries as functions of
the atomic number $A$
should be small for small $A$ (when
the Bethe-Heitler amplitude dominates); they should raise when DVCS amplitude becomes comparable with the
Bethe-Heitler amplitude.
As a  consequence, the asymmetries should have their maxima.
However
the position of the maxima is strongly model dependent:
the model (\ref{HMesParam}) predicts the maximum at moderate values of $A\propto 40-50$ for the beam-charge asymmetry,
 whereas the model~(\ref{OwensParam}) predicts it for $A>208$ (i.e. that for all nuclei the $A$-dependence
 of the beam-charge asymmetry is homogeneous).
\par
\begin{figure}[h]
\includegraphics[scale=0.35]{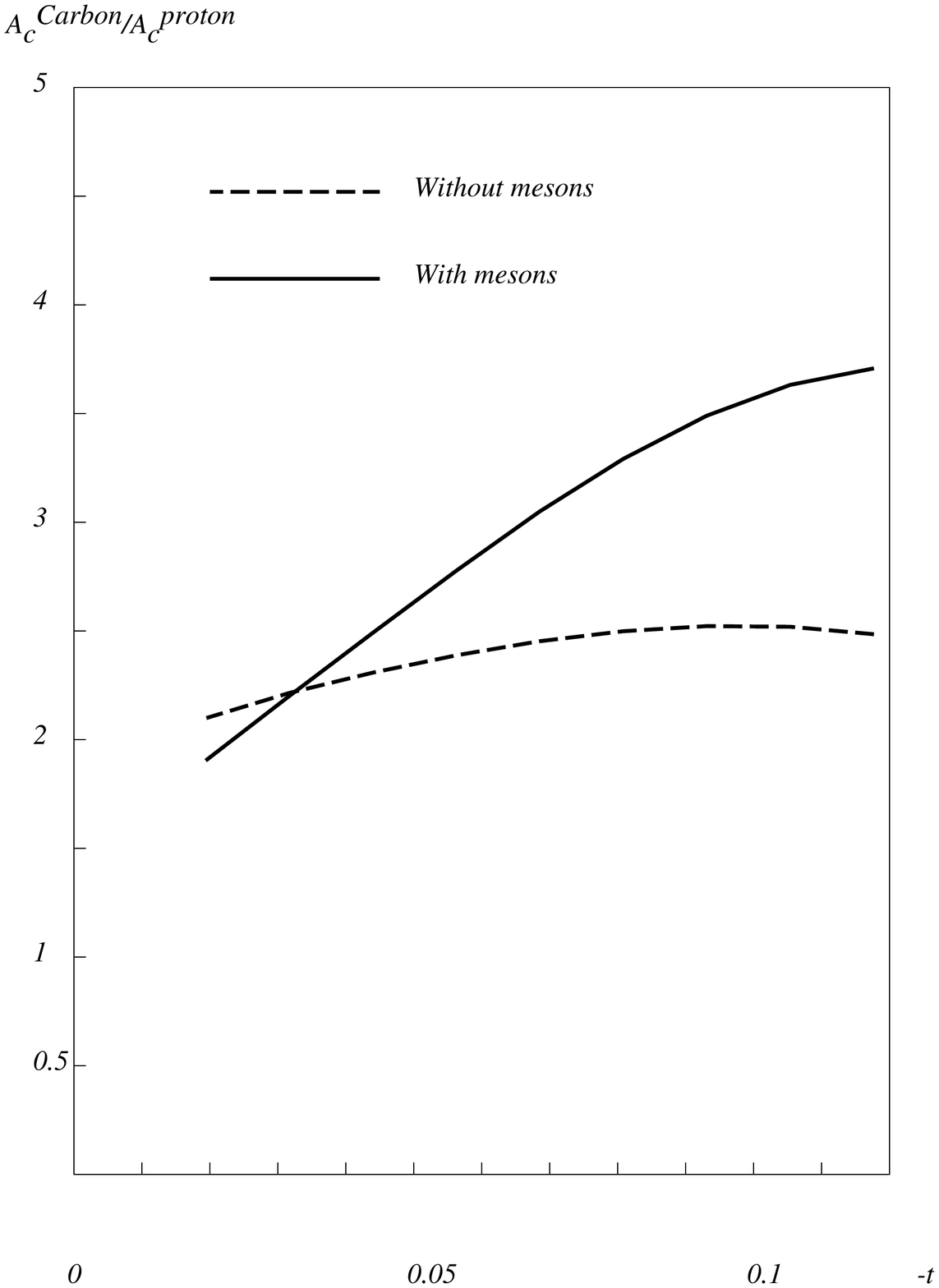}
\includegraphics[scale=0.35]{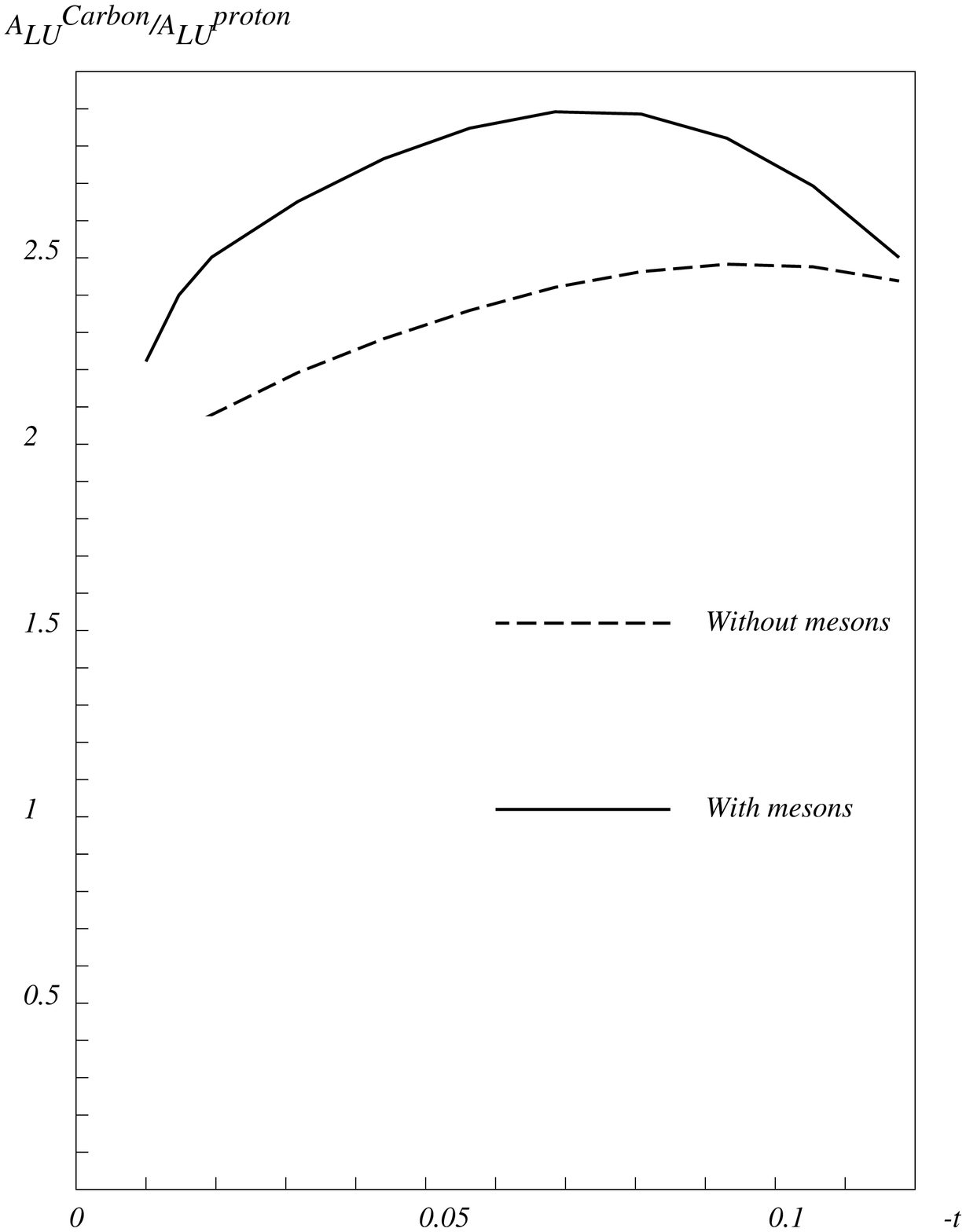}
\caption{\label{BCAt}
The ratio of the nuclear to the free proton beam-charge and beam-spin
asymmetries as functions of the
momentum transfer squared $t$ at fixed $\xi \approx 0.045$ (per nucleon)
for the C$^{12}$ nucleus. The results are presented for the full calculation
with  the parameterization (\ref{OwensParam})
(solid  curves)
and  for the calculation ignoring the meson contribution (dashed curves).}
\end{figure}

 In Fig.~\ref{BCAt} we present the $t$-dependence of the ratio of the nuclear to the
free proton beam-charge and beam-spin asymmetries at fixed $\xi \approx 0.045$ (per nucleon).
From Fig.~\ref{BCAt} one can see that the qualitative $t$-dependence of both asymmetries
does also change in the presence of mesons.
\par It is interesting to extrapolate our results to the point $t=0$ to eliminate the "pure kinematical"
$A$-dependence coming from the difference of the $\langle r^2\rangle$ in the nuclear formfactor $F_A(t)=Z e^{-\langle r^2\rangle |t|/6}$. The fit to data gives $|A_{DVCS\,A}/A_{DVCS\,N}|^2\propto A^{4.57\pm 0.17}$,
which is remarkably close to the value $d_A^2(0)\propto A^{4.52}$ obtained in Section~\ref{SectDterm}. In the asymmetries nuclear formfactors contract and we get for $t=0$ approximately the same power as for finite $t$. 
\begin{center}
\begin{table}[h]
\begin{tabular}{|l|r|r|r|r|r|r|}
\hline
Nucleus & $A^{cos}_{C\,\,A}/A^{cos}_{C\,\,N}$ & $A^{sin}_{LU\,\,A}/A^{sin}_{LU\,\,N}$ &
 $A^{cos}_{C\,\,A}/A^{cos}_{C\,\,N}$ & $A^{sin}_{LU\,\,A}/A^{sin}_{LU\,\,N}$&
  $A^{cos}_{C\,\,A}/A^{cos}_{C\,\,N}$ & $A^{sin}_{LU\,\,A}/A^{sin}_{LU\,\,N}$\\
\hline
$ ^{12}C   $  &2.45  &1.85 & 4.61  & 2.49 & 1.573 & 2.222\\
$ ^{16}O   $  &2.43  &1.83 & 5.41  & 2.33 & 1.905 & 2.270\\
$ ^{40}Ca  $  &2.38  &1.79 & 7.34  & 1.60 & 3.276 & 2.180\\
$ ^{90}Zr  $  &2.59  &1.93 & 6.80  & 0.81 & 4.879 & 2.104\\
$ ^{208}Pb $  &2.42  &1.07 & 6.12  & 0.31 & 6.077 & 0.998\\
\hline
\end{tabular}
\caption{\label{tabasym} The ratios of the
nuclear to the free proton
asymmetries for different nuclei.
The second and third columns correspond to the calculation without
the nuclear mesons; the fourth and fifth columns correspond to
the full calculation including the meson contribution with the
internal structure model given by (\ref{HMesParam}); the sixth and the
seventh columns correspond to the parameterization (\ref{OwensParam}).}
\end{table}
\end{center}

\par
It is interesting to compare our predictions for the DVCS asymmetries to the HERMES preliminary
result on Neon~\cite{Ellinghaus:2002zw}. Taking the ratio of the nuclear to the proton single-spin
asymmetries measured by HERMES, one readily finds
\bea
\frac{A^{sin}_{LU\,\,A}}{A^{sin}_{LU\,\,N}}=1.22 \pm 0.26 \,,
\eea
where we have added the proton and neon experimental errors in quadrature.
In the similar kinematics, the linear interpolation of our results from Table~{\ref{tabasym}}
gives
\bea
\frac{A^{sin}_{LU\,\,A}}{A^{sin}_{LU\,\,N}} \approx 2.1 \pm 0.2 \,.
\eea
A possible explanation of the discrepancy between the experimental result and the
theoretical prediction (a simple combinatoric counting also suggests the enhancement) was
suggested in~\cite{Guzey:2003jh}. The explanation consists in the observation
that, depending on $t$ and the experimental resolution,
the contribution of incoherent nuclear scattering might significantly reduce the
nuclear DVCS asymmetries.
\section{Conclusions}
\label{SectConclusion}
We performed a microscopic evaluation of nuclear GPDs for spin-0 nuclei
 in the framework of the Walecka model.
We found that the meson (non-nucleonic) degrees of freedom of the Walecka model play
a dramatic role in nuclear GPDs.

We observed a non-trivial $A$-dependence of the first moment of the nuclear $D$-term,
$d_A(0) \propto A^{2.26}$, which confirmed the prediction of M.~Polyakov made in the
framework of the nuclear liquid drop model~\cite{Polyakov:2002yz}.
This result demonstrated that contrary to the
 assumptions of the random phase approximation, mesonic degrees of freedom
dominate $d_A(0)$ and cannot be neglected.

Using the resulting nuclear GPDs, we studied the $A$ and $t$-dependence of the beam-charge
and beam-spin DVCS asymmetries in the HERMES kinematics. We found that due to the mesonic
degrees of freedom, the nuclear DVCS amplitude squared has a rapid $A$-dependence,
$|A_{DVCS}|^2 \propto A^{4.29}$, and becomes much larger than the Bethe-Heitler amplitude squared.
This dramatically affects the $A$-dependence of the asymmetries.
We found that $A^{cos}_{C\,\,A}/A^{cos}_{C\,\,N}\propto A^{0.5}$;
$A^{sin}_{LU\,\,A}/A^{sin}_{LU\,\,N}\propto A^{-0.03}$  for all $A$.
This behaviour should be compared to the case, when the meson degrees of freedom are neglected.
In this case, the DVCS asymmetries are virtually $A$-independent.

\section*{Acknowledgments}
\par We would like to thank K. Goeke, P. Pobylitsa, M. Polyakov,
M. Strikman and P. Schweitzer for valuable discussions and encouragement.
We also would like to thank K. Tsushima for providing us the code of TIMORA program
and discussions of the Walecka model.
This work is supported by the Sofia Kovalevskaya Program of the Alexander von Humboldt Foundation (V.G.)
and the Graduiertenkolleg 841 of DFG (M.S.).
\appendix*
\section{Derivation of the GPDs}
\label{SectApp1}
In this section we establish the formulas used for evaluation of GPDs in the functional integrals approach~\cite{Faddeev:1977rm}.
The generalized parton distributions are defined as
\bea
&&\nonumber H_\psi(x,\xi,t)=\half\int \frac{dz^-}{2\pi}e^{ix\bar P^+z^-}\langle P'|\bar \psi(\frac{-z}{2}n_-)\gamma_+\psi(\frac{z}{2}n_-)|P\rangle\,, \\
&&\nonumber H_\phi(x,\xi,t)=\frac{1}{2xP^+}\int \frac{dz^-}{2\pi}e^{ix\bar P^+z^-}\langle P'|\partial_+ \phi(-\frac{z}{2}n_-)\partial_+\phi(\frac{z}{2}n_-)|P\rangle\,, \\
&&\nonumber H_V(x,\xi,t)=\frac{1}{2xP^+}\int \frac{dz^-}{2\pi}e^{ix\bar P^+z^-}\langle P'| V_{+}^\alpha(-\frac{z}{2}n_-)V_{+\alpha}(\frac{z}{2}n_-)|P\rangle\\
&&\label{AGPDDef}+\frac{m_V^2}{2xP^+}\int \frac{dz^-}{2\pi}e^{ix\bar P^+z^-}\langle P'| V_{+}(-\frac{z}{2}n_-)V_{+}(\frac{z}{2}n_-)|P\rangle\,.
\eea
For the sake of brevity we denote $\Phi(x)=\{\phi(x),V(x),\bar\Psi(x),\Psi(x)\}$
and notice that in all three cases we have to evaluate the general matrix element
\bea
\langle P'|\Phi\left(\frac{z}{2}\right)\Phi\left(-\frac{z}{2}\right)|P\rangle=
\frac{\int D\phi(x) D\bar\Psi(x)D\Psi(x) D V(x) \Phi\left(\frac{z}{2}\right)\Phi\left(-\frac{z}{2}\right) e^{i\,S}}
{\int D\phi(x) D\bar\Psi(x)D\Psi(x) D V(x) e^{i\,S}  },\label{FaddInt}
\eea
which, up to inessential kinematical factors and integration over the light-cone separation $z$,
will give us the GPDs~(\ref{AGPDDef}).
\par
By definition we should integrate in (\ref{FaddInt}) over all configurations which
satisfy the asymptotic conditions~\cite{Faddeev:1977rm}
\bea
\lim_{t\rightarrow \pm \infty} \Phi(\vec x,t)\approx   \Phi^{P,P'}(\vec x,t)\,,
\eea
or in more explicit form
\bea
\lim_{t_1\rightarrow +\infty} \Phi(\vec x,t)\approx   \Phi_s(\vec x- \vec V' t_1)\,,&&
\lim_{t_2\rightarrow -\infty} \Phi(\vec x,t)\approx   \Phi_s(\vec x- \vec V t_2)\,.
\eea
We reparameterize the field configurations as
\bea
&&\Phi(x)\rightarrow \Phi_{s}(\vec x-\vec X(t))+\delta\Phi(\vec x-\vec X(t))\,,\nonumber\\
&&\int D\Phi\rightarrow \int D\delta\Phi D\vec X(t)\,,
\eea
where the integral over the zero mode - position of the soliton $\vec X(t)$- should be
taken exactly, while the integral over the shifts $\delta \Phi(x)$
may be evaluated in the saddle-point approximation. 
Notice that the standard mean field approximation
ignores all the loop corrections and thus implicitly uses the smallness of the coupling constant $g$.
In this case the evaluation of a simple path integral for the center of mass motion gives
\bea
&&\nonumber\label{CMIntegral}\int D \Phi(t)\Phi\left(\frac{z}{2}\right)\Phi\left(-\frac{z}{2}\right) e^{i\,S}\approx
e^{i\,S_{cl}}\int d\vec X_1\int d\vec X_2  \Phi\left(\frac{z}{2}-\vec X_1\right)\Phi\left(-\frac{z}{2}-\vec X_2\right)\times \\
&&\nonumber\times\int \prod_{t>z^0/2} d\vec X(t)e^{i\,M_s\,\dot{\vec X}(t)^2/2}
\int \prod_{t<-z^0/2}d\vec X(t)e^{i\,M_s\,\dot{\vec X}(t)^2/2}
\int \prod_{-z^0/2<t<z^0/2}d\vec X(t)e^{i\,M_s\,\dot{\vec X}(t)^2/2}\\
&&\sim\frac{const}{\sqrt{z_0}}\int d\vec X_1d\vec X_2\exp\left[\frac{iM_s (\vec X_1-\vec X_2)^2}{2z_0}\right]
\exp(i(\vec P'\vec X_1-\vec P\vec X_2))\Phi_s\left(\frac{z}{2}-\vec X_1\right)\Phi_s\left(-\frac{z}{2}-\vec X_2\right)
\eea
where $\vec X_1=\vec X(t=z^0/2); \vec X_2=\vec X(t=-z^0/2)$
and continuity of the path as well as proper boundary conditions
$$\vec X(t_1)=\vec q_1,\,\,\vec X(t_2)=\vec q_2,\,\,\lim_{t_1 \rightarrow \infty }\frac{\vec q_1}{t_1}=\vec V',
\,\,\lim_{t_2 \rightarrow -\infty }\frac{\vec q_2}{t_2}=\vec V$$ are implied;
the $const$ represents (divergent) kinematical factors which will contract with the same factors
in the denominator of (\ref{FaddInt}).
From the definition (\ref{AGPDDef}) we can see that the most essential are
small $z\sim 1/\bar P^+$. The term $const\exp\left[\frac{iM_s(\vec X_1-\vec X_2)^2}{2z_0}\right]/\sqrt{z_0}$
is not analytical at the point $z_0=0$ and is strongly suppressed for $z_0\neq 0$. Hence we may safely
replace it with its limit
$$
\lim_{z\rightarrow 0}\frac{1}{\sqrt{z_0}}\exp\left[\frac{iM_s (\vec X_1-\vec X_2)^2}{2z_0}\right] \sim \delta(\vec X_1-\vec X_2)
$$
\par Thus finally we arrive to the well-known result \cite{Rajaraman,Diakonov:1997vc}
\bea
\langle P'|\Phi\left(\frac{z}{2}\right)\Phi\left(-\frac{z}{2}\right)|P\rangle
\approx \frac{M_s}{2\pi}\int d^3 X \Phi_s\left(\frac{z}{2}-\vec X\right)\Phi_s\left(-\frac{z}{2}-\vec X\right)e^{i\vec \Delta \vec X}
\eea
which is valid in the leading order in the coupling constant: to evaluate the matrix element, one should replace all the fields
with the solutions of the Hartree-Fock equations centered at the point $\vec X$  and integrate over
the center of mass.

\end{document}